\documentclass[%
aip,
amsmath,amssymb,jcp,
groupedaddress,
preprint
]{revtex4-1}

\usepackage{float}

\usepackage[normalem]{ulem}

\usepackage{graphicx}
\usepackage{dcolumn}
\usepackage{bm}

\usepackage[utf8]{inputenc}
\usepackage[T1]{fontenc}
\usepackage{mathptmx}
\usepackage{hyperref}
\usepackage{etoolbox}
\usepackage{outlines}
\usepackage{color}
\usepackage{longtable}
\usepackage{array}
\usepackage{chngcntr}
\usepackage[version=4]{mhchem}
\usepackage[normalem]{ulem} 

\usepackage{amsmath}
\usepackage{amsfonts}
\usepackage{amssymb}
\newcolumntype{L}[1]{>{\raggedright\arraybackslash}p{#1}} 
\newcolumntype{C}[1]{>{\centering\arraybackslash}p{#1}}   
\newcolumntype{R}[1]{>{\raggedleft\arraybackslash}p{#1}}  

\usepackage{booktabs}
\usepackage{siunitx}
\usepackage{array,booktabs,longtable,amsmath}
\DeclareRobustCommand{\cation}{\mathrm{Na}}

\usepackage{soul}
\usepackage{xcolor}
\usepackage{hyperref}
\usepackage[nameinlink]{cleveref}

\definecolor{mycolor}{RGB}{255,165,0} 

\makeatletter
\def\@email#1#2{%
 \endgroup
 \patchcmd{\titleblock@produce}
  {\frontmatter@RRAPformat}
  {\frontmatter@RRAPformat{\produce@RRAP{*#1\href{mailto:#2}{#2}}}\frontmatter@RRAPformat}
  {}{}
}%
\makeatother
\begin{document}
\title{Thermal Transport Anomalies of Electrolyte Solutions in the Water Supercooled Regime: Signatures of the Liquid-Liquid Water Phase Transition}

\author{Guansen Zhao}
\affiliation{ 
Department of Chemistry, Molecular Sciences Research Hub, Imperial College London, London, W12 0BZ, United Kingdom 
}%

\author{Fernando Bresme$^*$}
\affiliation{ 
Department of Chemistry, Molecular Sciences Research Hub, Imperial College London, London, W12 0BZ, United Kingdom 
}%
\email[Corresponding author:]{f.bresme@imperial.ac.uk}

\date{\today}

\begin{abstract}
Water exhibits remarkable anomalies when supercooled, attributed to a hypothesized liquid-liquid phase transition (LLPT) between low-density (LDL) and high-density (HDL) liquid phases. Using non-equilibrium molecular dynamics simulations, we explore thermal transport and coupled effects in supercooled NaCl and LiCl solutions (1–4 m, 200–300 K). At 1 m, thermal conductivity exhibits a pronounced minimum near 220 K, coinciding with maxima in isothermal compressibility and minima in the speed of sound, both of which are signatures of critical fluctuations. The anomalies progressively diminish with increasing salt concentration and vanish at 4 m, suggesting suppression of the LLPT. The Soret coefficient exhibits a striking behavior. Initially thermophobic at high temperatures ($>$ 280 K), becoming thermophilic upon cooling, then reverting to thermophobic below 220 K. This behavior correlates with structural changes in the hydrogen-bond network of water. Specifically, we find that electrolyte solutions dominated by HDL structures, which are characterized by lower tetrahedral order, exhibit thermophobic behavior, whereas thermodynamic states dominated by LDL structures, with higher tetrahedral order, display thermophilic behavior. Furthermore, Seebeck coefficients exhibit sign reversals near 220–230 K, highlighting the thermoelectric sensitivity to structural transformations and temperature. These findings establish thermal transport as a sensitive probe of supercooled water, revealing that electrolyte solutions preserve the water's anomalies deep into the supercooled regime.

\end{abstract}

\maketitle

\section{Introduction}

Water, being the most abundant liquid on Earth, plays a central role in various natural and technological processes within the physical, chemical, and biological sciences.\cite{Franks2000} Unlike most liquids, water exhibits a density maximum and displays anomalous behaviors, such as increased isothermal compressibility and heat capacity when cooled.\cite{10.1093/acprof:oso/9780198570264.001.0001, Debenedetti_book1996} These anomalies become even more pronounced in the supercooled regime,\cite{speedy_isothermal_1976,angell_heat_1982,debenedetti_supercooled_2003} raising fundamental questions about the underlying mechanisms governing water's structure and dynamics.

Several hypotheses have been proposed to explain these anomalies, with one of the most compelling being the liquid-liquid phase transition (LLPT) hypothesis. This hypothesis suggests that there is a transition between a low-density liquid (LDL) phase, which exhibits strong tetrahedral order, and a high-density liquid (HDL) phase, characterized by weaker local order.\cite{poole_phase_1992}  This transition is believed to terminate at a liquid–liquid critical point. The LLPT hypothesis attributes the anomalous thermodynamic behavior of supercooled water to critical fluctuations occurring as the liquid approaches the second critical point.\cite{gallo_water_2016}  

Recent simulation studies employing empirical and {\it ab initio} trained machine learning-based models have provided increasing evidence supporting the LLPT hypothesis.\cite{palmer_metastable_2014,debenedetti_second_2020,donkor_beyond_2024,malek_liquid-liquid_2025,sciortino_constraints_2025} Furthermore, experimental investigations, although technically challenging due to the rapid crystallization of water, have also yielded tentative support for the existence of a second critical point.\cite{mishima_decompression-induced_1998,gallo_water_2016,kim_maxima_2017,kim_experimental_2020,amann-winkel_liquid-liquid_2023} 

The studies mentioned earlier clearly demonstrate the existence of thermodynamic anomalies in pure water. However, it is crucial to understand how these anomalies behave in aqueous solutions, both for fundamental research and practical applications. This understanding is important because, in both natural and technological contexts, water rarely exists on its own and usually acts as a solvent. In many cases, water can be more readily supercooled in solution than in its pure form.\cite{kanno_homogeneous_1977} Simulation studies of NaCl and LiCl aqueous solutions using various force fields support the existence of a liquid–liquid critical point (LLCP), which shifts to higher temperatures and lower pressures upon salt addition.\cite{corradini_route_2010,corradini_liquidliquid_2011,perin_phase_2023} Notably, the LLCP has been reported to disappear in concentrated LiCl solutions (2.0 m).\cite{perin_phase_2023} These findings motivate further investigation into how ion type and concentration influence the LLCP location, and whether these structural changes are connected to the anomalies in supercooled aqueous electrolyte solutions.

Previous investigations into the thermal transport properties of supercooled pure water have reported a minimum in thermal conductivity \cite{biddle_thermal_2013,kumar_thermal_2011,bresme_communication_2014,zhao_thermal_2024}. This minimum has been connected to various thermodynamic anomalies, such as maximum isothermal compressibility and minimum adiabatic speed of sound. These phenomena are believed to be associated with the hypothesized liquid-liquid critical point (LLCP).

The thermal conductivity quantifies the magnitude of the heat flux emerging from an imposed temperature gradient.  Thermal gradients in pure polar fluids
 induce the fluid polarization, an effect known as the thermal polarization (TP) effect\cite{bresme_water_2008,chapman_polarisation_2022}. In electrolyte solutions, thermal gradients give rise to thermodiffusion (Ludwig-Soret effect)\cite{Ludwig,Soret} and thermoelectricity (Seebeck effect)\cite{seebeck_ueber_1826}. These non-equilibrium effects\cite{degroot} involve coupling between mass and heat fluxes, resulting in the development of concentration gradients (Soret) in fluid mixtures or electrostatic potential differences (Seebeck) in metals, as well as in both aqueous and non-aqueous solutions.\cite{bonetti_thermoelectric_2015,bonetti_huge_2011,di_lecce_thermal_2018,nickel_water_2024} Indeed the Seebeck effect has gained increasing attention for enabling electric and electrochemical devices to harvest low-grade heat through temperature-induced voltages, with ionic systems offering promising energy conversion.


The Soret effect is an intriguing transport phenomenon,\cite{wiegand_thermal_2004,kohler_soret_2016} that is increasingly being used in practical applications as a powerful tool for assessing protein-ligand binding affinities\cite{wienken_protein-binding_2010, seidel_microscale_2013,niether_thermophoresis_2019}. However, we still lack a comprehensive theory to predict whether a solute will migrate towards hot (thermophilic) or cold (thermophobic) regions. Additionally, the Soret coefficient often exhibits an exotic behavior that poses challenges to microscopic interpretation. For instance, minima in the Soret coefficient have been observed in both experiments and simulations of electrolyte solutions, revealing a strong correlation between these minima and the overall thermodiffusion response.\cite{colombani_thermal_1999,di_lecce_computational_2017,rudani_analyzing_2025} Recent simulation studies have also demonstrated a relationship between thermophobicity in electrolyte solutions and the disruption of the tetrahedral structure of water.\cite{zhao_alkali_2025} This finding aligns with earlier conclusions drawn from experimental analyses of aqueous solutions.\cite{niether_thermophoresis_2019}


In this work, we employ non-equilibrium molecular dynamics (NEMD) simulations to explore the thermal transport properties of NaCl and LiCl aqueous solutions within the supercooled water regime. We investigate the thermophysical and structural properties in these supercooled states, with a particular focus on isothermal compressibility due to its relevance to the hypothesized liquid-liquid phase transition (LLPT). 

We demonstrate that the thermodiffusive response of aqueous alkali halides is quite complex, featuring multiple thermophobic and thermophilic transitions. This complexity is evident from the presence of minima in the Soret coefficient and a prevalent thermophobic response observed at both high and very low temperatures. In contrast, a thermophilic response occurs at intermediate temperatures. Furthermore, we identify a minimum in thermal conductivity at a concentration of 1 m, which disappears in highly concentrated solutions, 4 m. We also discuss the potential thermodynamic and structural origins of these observations.

\section{Methods}

Water molecules were modeled using the TIP4P/2005 model,\cite{abascal_general_2005} a non-polarizable, rigid water model widely recognized for accurately reproducing the properties of supercooled water and various ice polymorphs.\cite{vega_simulating_2011,vega_what_2008} Ions were simulated using the Madrid 2019 force field,\cite{zeron_force_2019}, which is specifically parameterized for compatibility with TIP4P/2005 and employs scaled ionic charges to capture the thermodynamic and structural properties of electrolyte solutions. This force field has been shown to reliably reproduce experimental thermal transport behavior of NaCl and LiCl under ambient conditions\cite{bresme_thermal_2024}, and it has also been applied in simulations of supercooled electrolyte solutions.\cite{perin_phase_2023, sedano_isothermal_2024}

\begin{figure}[ht!]
    \centering
    \includegraphics[width=1\linewidth]{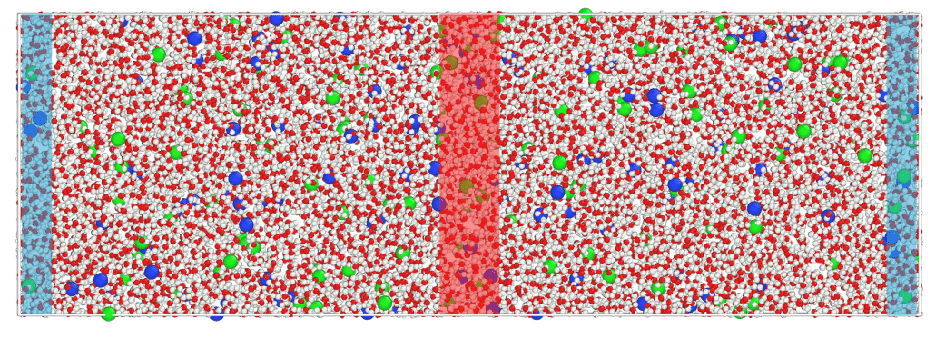}
    \caption{Snapshot of the simulation cell used in the NEMD simulation of a 1 m NaCl aqueous solution, visualized using OVITO.\cite{stukowski_visualization_2009} Red and white spheres indicate oxygen and hydrogen atoms, while blue and green spheres represent Na$^+$ and Cl$^-$ ions, respectively. The thermostatting regions are highlighted in red (hot, at the center of the box) and blue (cold, at the edges).}
    \label{fig:sim_box}
\end{figure}

We used boundary-driven non-equilibrium molecular dynamics (NEMD) simulations, performed with LAMMPS (version from 29 Aug 2024),\cite{plimpton_fast_1995} to investigate the thermal transport and the induced coupling effects of the alkali halide aqueous solutions. We applied thermostats to designated regions at the center and edges of the box, maintaining predefined hot ($T_\text{hot}$) and cold ($T_\text{cold}$) temperatures (see Figure \ref{fig:sim_box}). The thermostatting regions were 8~\AA \ thick, and the simulation box was configured with dimensions in the ratio \( \{L_x:L_y:L_z\} = \{1:1:3\} \), where \( L_x \) was approximately \( 40\,~\text{\AA} \), depending on the density. 

Langevin thermostats\cite{schneider_molecular-dynamics_1978,brunger_stochastic_1984} were employed to control the temperature of molecules and ions at each timestep (1 fs) with a coupling constant of 500 fs. We also removed the center of mass momentum of the entire system at every timestep. Electrostatic interactions were computed using the Particle-Particle Particle-Mesh (PPPM) method, \cite{Hockney1989} with a relative error of 10$^{-5}$. The PPPM method was likewise employed to handle long-range dispersion interactions,\cite{isele-holder_reconsidering_2013,isele-holder_development_2012} with accuracy thresholds set to 10$^{-4}$ kcal/(mol \AA) for the real space component and 2$\times$10$^{-4}$ kcal/(mol \AA) for the k-space component. 

A representative simulation box contained 6663 water molecules modeled using the TIP4P/2005 potential,\cite{abascal_general_2005} along with 120, 240, or 480 ion pairs, corresponding to concentrations of 1, 2, and 4 m, respectively. Ion-ion and ion-water interactions were described using the Madrid-2019 force field.\cite{zeron_force_2019} Prior to the NEMD runs, the system was pre-equilibrated for 5 ns in the isothermal-isobaric (NPT) ensemble at a pressure of 1 bar and a temperature set to the average $T_\text{hot}$ and $T_\text{cold}$. The equilibration process employed a Nos\'e-Hoover thermostat and barostat with relaxation times of 100 fs and 1000 fs, respectively.\cite{nose_molecular_1984,hoover_canonical_1985} Then the thermostats were activated, and the NEMD simulations were conducted under constant volume conditions.

We carried out long simulations for supercooled systems. The simulations lasted at least 200 ns, discarding the initial 150 ns to ensure the system had reached a steady state (see Figure \ref{fig:time-evo} in Appendix \ref{sec:time-evo}). Following this initialization period, we computed the local temperature and concentration profiles by averaging over 200 spatial bins along the $z$-axis, corresponding to a bin width of approximately \( 0.6\,~\text{\AA} \). These profiles were further averaged over a production run of at least 50 ns across 10 independent replicas, each initialized with a different random seed for the thermostats, yielding a total sampling time of 0.5 $\mu$s. The trajectory from at least the final 10 ns of the production run was saved every 1000 steps, resulting in a total of $10^5$ configurations used to compute tetrahedral order parameters and electrostatic potential profiles.

The thermal conductivity, $\lambda(z)$, along the direction of the local thermal gradient $\nabla T(z)$, was computed with Fourier's Law,
\begin{equation}
\lambda(z) = - \frac{J_q}{\nabla T(z)}  
\label{eqn: fourier}
\end{equation}
Here, $J_q$ represents the heat flux, which was determined using the continuity equation: 
\begin{equation}
  J_{q} = \frac{\dot{Q}}{2A}, 
\label{eqn: heat-flux}
\end{equation}
where $\dot{Q}$ denotes the rate of energy exchanged at the hot and cold thermostats, $A$ represents the cross-sectional area of the simulation box perpendicular to the direction of heat flow, and the factor of 2 accounts for the presence of the two heat fluxes within the simulation cell. Our approach ensures excellent energy conservation, with equal amounts of energy transferred at the cold and hot thermostats within computational accuracy (see Figure \ref{fig:energy_conserv} in Appendix \ref{sec:energy_conserv}).

The Soret coefficient $s_T$ was determined from the local molar fraction profiles obtained in NEMD production runs:
\begin{equation}
   s_T(T)= -\frac{1}{x_1(T) (1-x_1(T))}\left ( \frac{\nabla x_1(T)}{\nabla T(T)} \right )_{J_1 =0}= -\frac{1}{b(T)} \left ( \frac{db(T)}{dT} \right )_{J_1 =0} , 
\label{eqn:Soret}
\end{equation}
where $x_1(T)$ denotes the molar fraction of salt at the local temperature \(T\), respectively, and \(b(T)\) is the corresponding solute molality. 
Typical local concentration profiles used for the Soret coefficient calculations are shown in Appendix \ref{sec:concentration-T} (see Figure 10).

We also calculated the tetrahedral order parameter \( \xi \) to investigate structural changes in the supercooled electrolyte solutions, using the following expression:\cite{errington_relationship_2001}
\begin{equation}
\xi = 1 - \frac{3}{8} \sum_{j=1}^{3} \sum_{k=j+1}^{4} \left( \cos \psi_{jk} + \frac{1}{3} \right)^2
\label{eqn:order_param}
\end{equation}
where \( \psi_{jk} \) is the angle between the lines connecting a central water molecule and its four nearest oxygen neighbors.

To investigate the thermoelectric effect in electrolyte solutions, we calculated the charge density $\rho_q(z)$,
\begin{equation}
  \rho_q(z) = \frac{1}{A} \Biggl \langle  \sum_{i=1}^{N_{charges}} q_{i} \delta (z-z_{i})  \Biggr \rangle  , 
\label{eqn: charge-density}
\end{equation}
where the angle brackets denote a time-averaged binning of the sum over all charged sites. In this expression, \( q_{i} \) is the charge of site \( i \), and \( z_{i} \) is its coordinate along the \( z \)-axis. In the system studied, the charged sites comprise the hydrogen (\(q_{H} = 0.5564e\)) and dummy atom (\(q_{M} = -1.1128e\)) of the TIP4P/2005 water model.\cite{abascal_general_2005} Additional contributions come from the scaled charges of the ions, modeled using the Madrid-2019 potential: (\(q_{Na^+} =q_{Li^+} = 0.85e\)) and (\(q_{Cl^-} = -0.85e\)).\cite{zeron_force_2019} The electric field was obtained by integrating the charge density:
\begin{equation}
    E_{z}(z) = \frac{1}{\varepsilon_{0}} \int_{0}^{z} \rho_q (z') \,dz', \label{eqn: Efield}
\end{equation}
where $\varepsilon_{0}$ is the vacuum permittivity. To quantify the strength of the thermoelectricity response, we calculated the Seebeck coefficient $S_E$,
\begin{equation}
    S_{E}(z) = \frac{E_z(z)}{\nabla T(z)}, 
\label{eqn:Seebeck}
\end{equation}
which represents the ratio of the $z$ component of the electrostatic field to the thermal gradient. The detailed procedure for calculating the Seebeck coefficient from the local electrostatic potential profiles is provided in Appendix \ref{sec:Potential-T}.

To investigate the thermophysical properties of the electrolyte solutions, we performed equilibrium molecular dynamics simulations in the isothermal–isobaric (NPT) ensemble using GROMACS (vs. Jan 2024).\cite{Gromacs1993, GROMACS2024} Each system consisted of a cubic simulation box containing 926 water molecules and 17, 34, or 67 ion pairs, corresponding to 1.0, 2.0, and 4.0 m concentrations, respectively. The production runs were carried out for at least 200 ns with a timestep of 1 fs to ensure adequate sampling. The fluctuation-derived thermodynamic properties, including the isobaric thermal expansion coefficient (\(\alpha\)), isothermal compressibility (\( \kappa_T \)), the isobaric (\( C_P \)) and isochoric (\( C_V \)) heat capacities, were computed using the following fluctuation relations:\cite{Allen&Tildesley1987}
\(
\langle \delta V^2 \rangle_{NPT} = V k_B T \kappa_T, \;
C_P = \left < \delta H^2 \right> / k_BT^2,  \;
C_V = C_P - V T \alpha^2/\kappa_T.
\)
In these expressions, \( V \) is the system volume, \( H \) is the enthalpy, and \( k_B \) is the Boltzmann constant. The fluctuation \( \delta \mathcal{P}  = \mathcal{P} - \mathcal{P}_{ens}\) denotes the deviation of a thermodynamic property \( \mathcal{P} \) from its ensemble average \( \mathcal{P}_{\text{ens}} \) in the NPT ensemble. The isentropic speed of sound was then calculated from these fluctuation properties along with the mass density \( \rho \):
\begin{equation}
    c_{s} = \sqrt{\frac{C_{P}}{C_{V}} \frac{1}{\rho \kappa_T}}.
    \label{eqn:cs}
\end{equation}
\noindent

\section{Results}

\begin{figure*}[ht!]
    \centering
    \includegraphics[width=1\linewidth]{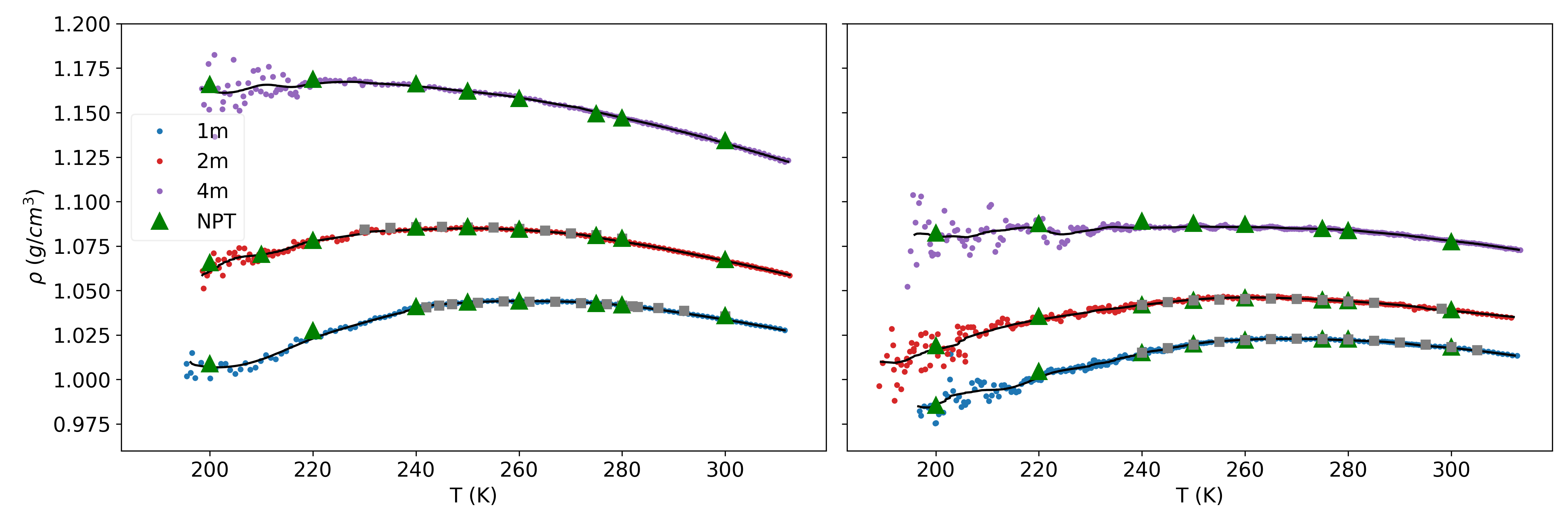}
    \caption{Equations of state obtained from equilibrium NPT simulations at 1 bar are represented by green triangles. Data from NEMD simulations are shown in blue, red, and purple, corresponding to 1, 2, and 4 m concentrations, respectively. The black curves indicate a running average over 10 consecutive points. Gray squares denote reference simulation data from Ref~\cite{sedano_maximum_2022}. The left panel presents results for NaCl solutions, while the right panel corresponds to LiCl solutions. Tables~\ref{table:NEMD-data} and~\ref{table:NPT-data} in the Appendix provide the numerical results for the NEMD and NPT simulations, respectively.}
    \label{fig:eos}
\end{figure*}

Figure \ref{fig:eos} illustrates the equations of state for NaCl and LiCl aqueous solutions at concentrations of 1, 2, and 4 molal (m), calculated along the standard pressure isobar (1 atm) using local temperatures and densities obtained from Non-Equilibrium Molecular Dynamics (NEMD) simulations. The results demonstrate a strong agreement with our own equilibrium NPT simulation data, as well as with previous molecular dynamics simulations \cite{sedano_maximum_2022}. This consistency between our equilibrium and non-equilibrium results, along with earlier studies, reinforces the validity of the simulation conditions applied in our NEMD computations. Our results indicate that the systems achieve local equilibrium under thermal gradients of approximately 1 K/\AA. In line with findings reported in earlier studies, our results show that the addition of salt causes a shift in the temperature of the density maximum (TMD) to lower temperatures. As discussed by \citet{sedano_maximum_2022}, the predictions made using the TIP4P/2005 - Madrid-2019 force field align closely with the experimental TMDs of LiCl and NaCl within the 0-2 m concentration range.

\begin{figure*}[ht!]
    \centering
    \includegraphics[width=1\linewidth]{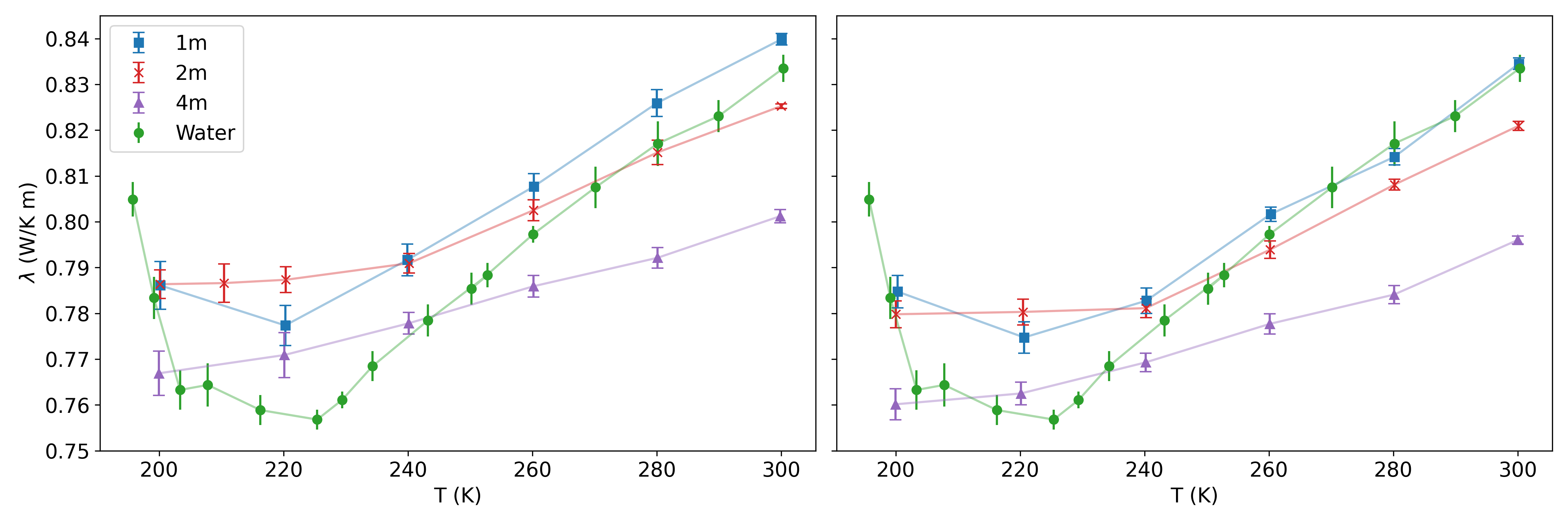}
    \caption{Thermal conductivity of NaCl (left) and LiCl (right) solutions at various concentrations over the 200–300 K temperature range. Green dots indicate thermal conductivity values for pure water reported in a previous study.\cite{zhao_thermal_2024} Lines connecting the data points are included as a visual guide.}
    \label{fig:therm_cond}
\end{figure*}

Figure \ref{fig:therm_cond} shows the thermal conductivities of NaCl and LiCl aqueous solutions as a function of temperature within the range of 200–300 K, at concentrations varying from 1 to 4 m. The thermal conductivity of these solutions decreases with temperature. However, for salt concentration 1 m, a minimum in the thermal conductivity is observed at lower temperatures, around 220 K. The results also show that the thermal conductivity decreases as the density of the fluid increases in a wide temperature range, indicating an anomalous behavior in the thermal response of the solution. Advancing the discussion below, this reduction is associated with a decrease in the speed of sound in the solution throughout most of the supercooled region examined here.

The thermal conductivity of NaCl and LiCl solutions is very similar across the studied temperature and concentration ranges. This behavior is consistent with previous simulation studies, which reported that the thermal conductivity of alkali halide solutions is more strongly influenced by anion mass and composition than by cation type. \cite{bresme_thermal_2024} At 300 K, both 1 m NaCl and LiCl solutions exhibit conductivities close to that of pure water, while at higher concentrations a pronounced reduction is observed. 

In both NaCl and LiCl solutions at 1 m concentration, the thermal conductivity exhibits a local minimum near 220 K, similar to what is observed in pure TIP4P/2005 water.\cite{zhao_thermal_2024} However, this minimum disappears at higher concentrations (2 m and 4 m) indicating that salt addition suppresses the thermal conductivity anomaly.

\begin{figure*}[ht!]
    \centering
    \includegraphics[width=1\linewidth]{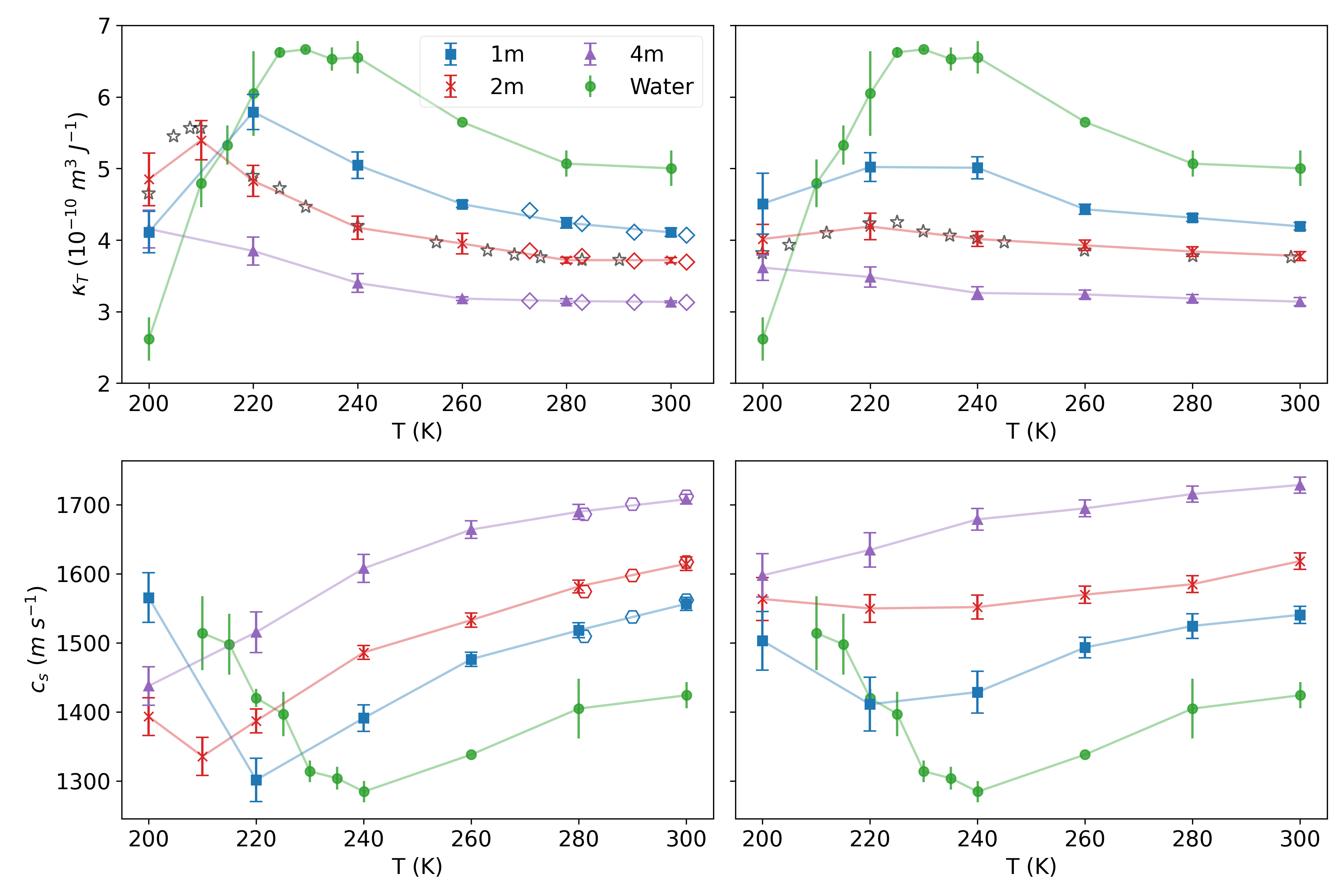}
    \caption{Isothermal compressibility (top panels) and speed of sound (bottom panels) of 1, 2, and 4 m NaCl (left) and LiCl (right) solutions as a function of temperature. Green dots show data for pure TIP4P/2005 water results from Ref~\cite{zhao_thermal_2024}. In the top left panel, open diamonds are experimental values for NaCl from Ref~\cite{rogers_volumetric_1982}. In both top panels, open star markers denote simulated isothermal compressibilities for 2 m NaCl and LiCl from reference \cite{sedano_isothermal_2024}. In the bottom left panel, open hexagons represent the speed of sound of NaCl from Ref~\cite{kleis_dependence_1990}, computed from the correlation equations reported in that work based on experimental measurements.}
    \label{fig:fluct}
\end{figure*}

The minimum in thermal conductivity suggests an underlying thermodynamic origin. To understand this connection, we examine the isothermal compressibility, which is expected to influence thermal transport due to its relationship to the speed of sound.
  The compressibility was computed using equilibrium simulations conducted in the NPT ensemble and was used to calculate the isentropic (adiabatic) speed of sound. Our findings on isothermal compressibility are in good agreement with both experimental data and previous simulation results, under both ambient and supercooled conditions \cite{rogers_volumetric_1982, sedano_isothermal_2024, bresme_thermal_2024} (see Figure \ref{fig:fluct}).

The isothermal compressibility of the NaCl and LiCl 1 m and 2 m solutions exhibits prominent maxima, resembling the behavior observed for pure water in earlier studies \cite{gallo_water_2016}. However, the general trend indicates that these maxima shift to lower temperatures and decrease in magnitude. This behavior is consistent with the findings reported in previous works \cite{perin_phase_2023, sedano_isothermal_2024}, and it has significant implications for the behavior of the speed of sound (see Figure \ref{fig:fluct}). The calculated speed of sound for NaCl aligns well with available experimental data \cite{kleis_dependence_1990}. It decreases with temperature in the moderate undercooling regime (T \(>\) 240 K) and in our simulations it features minima for the NaCl and LiCl 1 m solutions at temperatures corresponding to the points where compressibility exhibits a maximum. These minima coincide with the region where a minimum in thermal conductivity is also observed (see Figure \ref{fig:therm_cond}). We note that the influence of the speed of sound on thermal conductivity was previously noted by Bridgman in his empirical equation \cite{bridgman_thermal_1923}.

At higher concentrations, specifically 4 m, the speed of sound does not feature any distinct minima. The lack of clear minima within the temperature range studied may be attributed to a shift in the temperature of these extrema to values lower than those included in our investigation, as well as a further attenuation of the anomalies in more concentrated solutions. These findings collectively support the view that thermal transport in supercooled alkali halide solutions remains influenced by underlying thermodynamic anomalies. This interpretation suggests that the minimum in thermal conductivity is aligned with the presence of a liquid-liquid critical point (LLCP), which is evident in the 1 m solution. However, at higher concentrations, these anomalies are significantly diminished or no longer detectable, consistent with previous research indicating that the LLCP is suppressed or completely disappears in concentrated lithium chloride (LiCl) solutions.\cite{perin_phase_2023}

Our analysis of the speed of sound indicates a correlation with the lack of minima in the thermal conductivity of highly concentrated solutions. This suggests that the reduction in thermal conductivity observed across all salts and concentrations, along with the minima found in 1 m solutions, is related to the temperature dependence of the speed of sound. Ultimately, this behavior arises from the temperature dependence of the isothermal compressibility of the solutions in the supercooled regime.

\begin{figure*}
    \centering
    \includegraphics[width=1\linewidth]{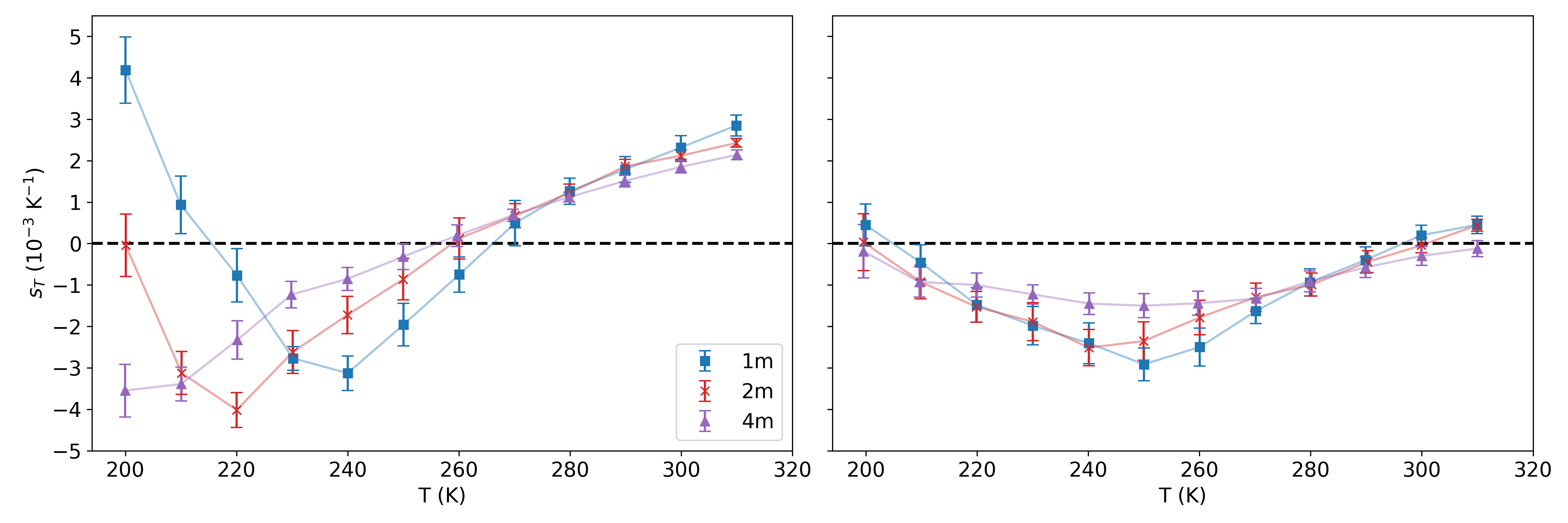}
    \caption{Soret coefficients for NaCl (left) and LiCl (right) solutions at various concentrations over the 200–300 K temperature range. The horizontal dashed black line represents a Soret coefficient of zero. }
    \label{fig:soret}
\end{figure*}

Having established the thermodynamic signatures of structural transitions, we now examine thermodiffusive transport, which provides insight into the relationship between thermal gradients and molecular mobility. In Figure \ref{fig:soret}, we present the Soret coefficients as a function of temperature for aqueous NaCl and LiCl solutions at various concentrations. At 300 K, our results align with previous simulation studies, accurately reflecting both the magnitude and temperature dependence of the Soret effect.\cite{bresme_thermal_2024} Notably, we observe a transition in the Soret coefficient from thermophobic behavior (at high temperatures) to thermophilic behavior (at low temperatures). This transition occurs for LiCl at approximately 300 K and for NaCl around 260 K, regardless of the concentration studied. This sign reversal has also been documented in experimental work and simulations using different force fields.\cite{gaeta_nonisothermal_1982,romer_alkali_2013,lee_non-monotonic_2024,di_lecce_role_2017,di_lecce_soret_2019}

A notable feature is the crossing of the Soret coefficient curves for different concentrations near 280 K. Above this temperature, the Soret coefficient decreases with increasing concentration, whereas below it, the trend is reversed. A similar crossing behavior was observed in experiments at approximately 315 K,\cite{romer_alkali_2013} as well as in simulations of LiCl solutions using the SPC/E water model, where the crossing occurred around 260–280 K.\cite{di_lecce_role_2017,di_lecce_computational_2017} This temperature interval is close to the value obtained here. These previous investigations attributed the concentration dependence to competing mechanisms: at high temperatures, entropy predominates, while at low temperatures, stronger hydrogen bonding enhances ion–water interactions, leading to an accumulation of salt in the heated region. Our results replicate these qualitative trends and extend them into the supercooled regime.

In the deeply supercooled regime, for both NaCl and LiCl (1 and 2 m), the Soret coefficient exhibits a distinct local minimum between 220 and 250 K (see Figure \ref{fig:soret}). This minimum shifts to lower temperatures as the concentration increases from 1 to 2 m. At a concentration of 4 m, NaCl does not exhibit any evidence of a minimum within the studied temperature range, while for LiCl, the minimum is not so clearly defined as compared with lower concentrations.

When comparing NaCl to LiCl, the Soret coefficient for NaCl displays a more significant change over a broader temperature range upon cooling, developing into a sharper minimum. Additionally, at temperatures close to 200 K, a second sign reversal occurs in the Soret coefficients of the 1 and 2 m solutions of both NaCl and LiCl. This transition is especially pronounced for the 1 m NaCl solution, where the Soret coefficient rises above \( 4 \times 10^{-3}\,\mathrm{K^{-1}} \) at 200 K, indicating a strong thermophobic response. Generally, our results suggest that supercooled solutions become less thermophilic with increasing concentration in the temperature range of 230-280 K and less thermophobic for temperatures above 280 K. This indicates an overall weakening of the thermodiffusion response as salt concentration increases.

\begin{figure*}[ht!]
    \centering
    \includegraphics[width=1\linewidth]{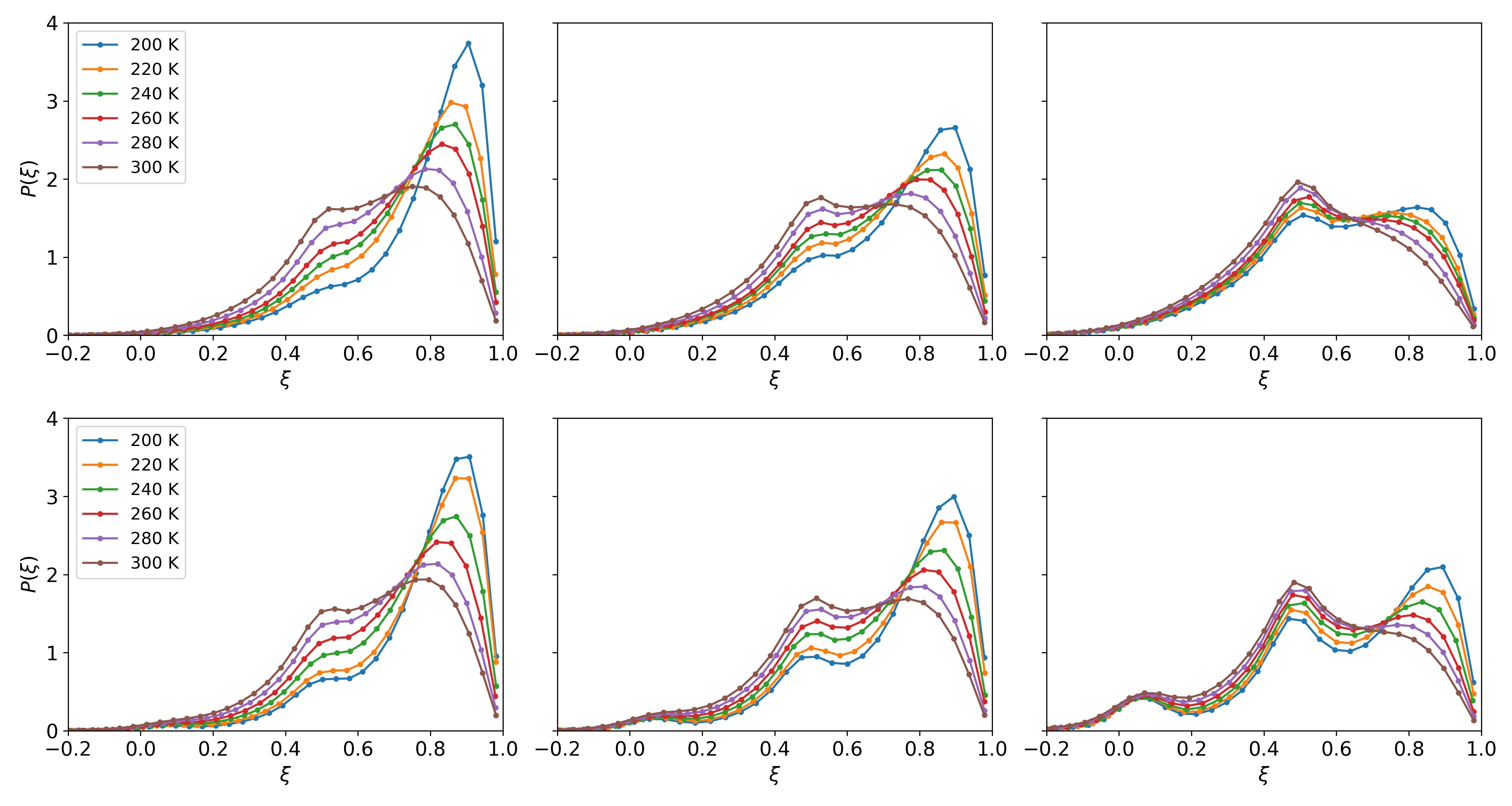}
    \caption{Probability density distribution of the tetrahedral order parameter \(\xi\) for NaCl (top row) and LiCl (bottom row) aqueous solutions at temperatures from 200 K to 300 K. Columns from left to right correspond to concentrations of 1, 2, and 4 m, respectively.}
    \label{fig:histogram}
\end{figure*}

The complex behaviour of the Soret coefficient suggests changes in the hydrogen-bond network of water. To explore the structural origins of the observed thermodiffusion behavior, we examine the tetrahedral order parameter. Figure \ref{fig:histogram} presents the distribution of the tetrahedral order parameter. The emergence of a bimodal distribution is consistent with previous findings in supercooled pure water, reflecting the coexistence of distinct local environments.\cite{errington_relationship_2001, kuo_tetrahedral_2021} In particular, the two dominant peaks correspond to more tetrahedrally ordered configurations associated with a low-density liquid–like (LDL) structures $(\xi \approx 0.8{-}1)$ and less ordered, high-density liquid-like (HDL) arrangements $(\xi \approx 0.4{-}0.6)$. A comparison across temperatures and concentrations indicates a tendency toward stronger tetrahedral ordering at lower temperatures and lower salt concentrations. 

Interestingly, for all the LiCl solutions, an additional peak emerges at $\xi < 0.2$,  indicating the existence of highly disordered local environments.  This multimodal distribution is absent in the case of NaCl, suggesting the emergence of heterogeneous structures in LiCl, which lead to the distortion of the water tetrahedral structure.

Due to its small radius, lithium produces a strong electrostatic field and is solvated by a stable hydration shell consisting of four to five water molecules, which involves a high heat of hydration~\cite{mahler_study_2012,mason_neutron_2015}. Both classical simulations and {\it ab initio} computations predict a solvation shell made up of four water molecules arranged in a tetrahedral motif.\cite{zeron_force_2019,rempe_hydration_2000} Experiments also indicate that the hydration number is relatively independent of ion concentration.\cite{mason_neutron_2015} These findings align with our calculations of the lithium-oxygen (Li-O) coordination number (CN) (see Table \ref{table:NPT-data} in Appendix \ref{sec:NPT-table}). Our simulations demonstrate that the CN is fairly consistent across different temperatures in the supercooled regime, which corroborates recent investigations \cite{perin_phase_2023}. In contrast, the coordination numbers for chloride-oxygen (Cl-O) and sodium-oxygen (Na-O) exhibit greater variability with respect to both temperature and concentration. Therefore, the observed increase in the height of the maximum across the 1-4 molality range (see Figure \ref{fig:histogram}) is likely reflecting the larger presence of [Li(H$_2$O)$_4$]$^+$ structures in the solution as the concentration increases.



We have analyzed the probability distributions in Figure \ref{fig:histogram} to quantify the relative fractions of HDL structures and their correlation with the Soret coefficients. The numerical data are listed in Table \ref{table:order_param}, and the detailed procedure for this calculation is provided in Appendix \ref{sec:Tetrahedral}. Between 240 and 300 K, the Soret coefficient for both NaCl and LiCl at all studied concentrations becomes more negative, {\it i.e} the solution becomes more thermophilic, with the ions migrating preferentially towards the hot region.  This observation is consistent with previous discussions that greater water ordering enhances the thermophobic character of aqueous solutions.\cite{niether_thermophoresis_2019,zhao_alkali_2025} At around 240 K, where the 1 m and 2 m NaCl and LiCl solutions exhibit local minima in the Soret coefficient, the calculated area fraction of the HDL structure is \(\sim\) 27\%, indicating that thermophilicity is observed in general for larger populations of LDL the structure, where the coordination of water molecules with other water molecules features a stronger orientational structure and lower coordination. Indeed, we find that over 70\% of the systems investigated in this work have an HDL area fraction below 30\%. Similarly, we find that all the systems with a high content of HDL, \(> 65\) \% have positive Soret coefficients, and are thermophobic. 

As we approach the supercooled regime, we observe a rapid increase in the Soret coefficient. It becomes strongly positive, \(\sim +4 \times 10^{-3} \) K\(^{-1}\), for the 1 M NaCl solution and slightly positive, \(\sim +0.5 \times 10^{-3} \) K\(^{-1}\) for the LiCl solution. These results suggest the presence of an additional competing mechanism influencing thermodiffusion in the deeply supercooled regime. It is possible that the order parameter we are examining is not sufficiently sensitive to capture other structural changes occurring in the supercooled aqueous solutions.

\begin{table}
\begin{ruledtabular}
\begin{tabular}{|c|ccc|ccc|ccc|}
& \multicolumn{3}{c|}{1 m} & \multicolumn{3}{c|}{2 m} & \multicolumn{3}{c|}{4 m} \\
\hline
T & $\overline{\xi}$ & HDL & $s_T$ & $\overline{\xi}$ & HDL & $s_T$ & $\overline{\xi}$ & HDL & $s_T$ \\
$[K]$ &  & $[\%]$ & $[10^{-3}\,\mathrm{K}^{-1}]$ & & $[\%]$ & $[10^{-3}\,\mathrm{K}^{-1}]$ &  & $[\%]$ & $[10^{-3}\,\mathrm{K}^{-1}]$ \\

\hline
\multicolumn{10}{|c|}{NaCl} \\
\hline
200 & 0.773 & 16.5 &  4.187(0.8) & 0.708 & 24.2 & -0.040(0.8) & 0.608 & 48.5 & -3.549(0.6) \\
220 & 0.738 & 21.1 & -0.767(0.6) & 0.683 & \textbf{27.1} & \textbf{-4.018(0.4)} & 0.592 & 48.8 & -2.325(0.5) \\
240 & 0.707 & \textbf{24.7} & \textbf{-3.129(0.4)} & 0.661 & 31.6 & -1.726(0.5) & 0.578 & 54.9 & -0.855(0.3) \\
260 & 0.683 & 28.9 & -0.747(0.4) & 0.637 & 33.3 &  0.122(0.5) & 0.564 & 68.1 &  0.193(0.3) \\
280 & 0.654 & 35.3 &  1.266(0.3) & 0.615 & 40.3 &  1.221(0.2) & 0.549 & 72.1 &  1.122(0.1) \\
300 & 0.628 & 34.3 &  2.316(0.3) & 0.594 & 51.4 &  2.120(0.1) & 0.534 & 75.5 &  1.850(0.1) \\
\hline
\multicolumn{10}{|c|}{LiCl} \\
\hline
200 & 0.766 & 15.7 &  0.449(0.5) & 0.719 & 22.4 &  0.035(0.7) & 0.625 & 36.9 & -0.186(0.6) \\
220 & 0.750 & 18.3 & -1.479(0.4) & 0.700 & 23.8 & -1.529(0.4) & 0.606 & 40.9 & -1.000(0.3) \\
240 & 0.717 & 22.9 & -2.407(0.5) & 0.672 & \textbf{27.2} & \textbf{-2.513(0.4)} & 0.586 & 45.5 & -1.450(0.3) \\
260 & 0.688 & \textbf{27.6} & \textbf{-2.499(0.5)} & 0.646 & 33.9 & -1.785(0.4) & 0.565 & 49.4 & -1.442(0.3) \\
280 & 0.658 & 32.8 & -0.937(0.3) & 0.620 & 35.1 & -0.991(0.3) & 0.545 & 57.4 & -0.914(0.2) \\
300 & 0.632 & 32.8 &  0.199(0.2) & 0.597 & 42.6 & -0.042(0.2) & 0.524 & 62.3 & -0.297(0.2) \\
\end{tabular}
\end{ruledtabular}
\caption{Average tetrahedral order parameter $\overline{\xi}$, percentage of HDL area fraction (HDL), and Soret coefficient $s_T$ for NaCl (top) and LiCl (bottom) aqueous solutions at 1, 2, and 4 m from 200–300 K temperature. $\overline{\xi}$ and HDL area fraction were obtained from tetrahedral order parameter distributions. The Soret coefficients are from the NEMD simulations with standard deviations from replicas reported in parentheses. Bold entries indicate the local minima of $s_T$ and matched HDL\% for the correlation analysis.}
\label{table:order_param}
\end{table}

\begin{figure*}
    \centering
    \includegraphics[width=1\linewidth]{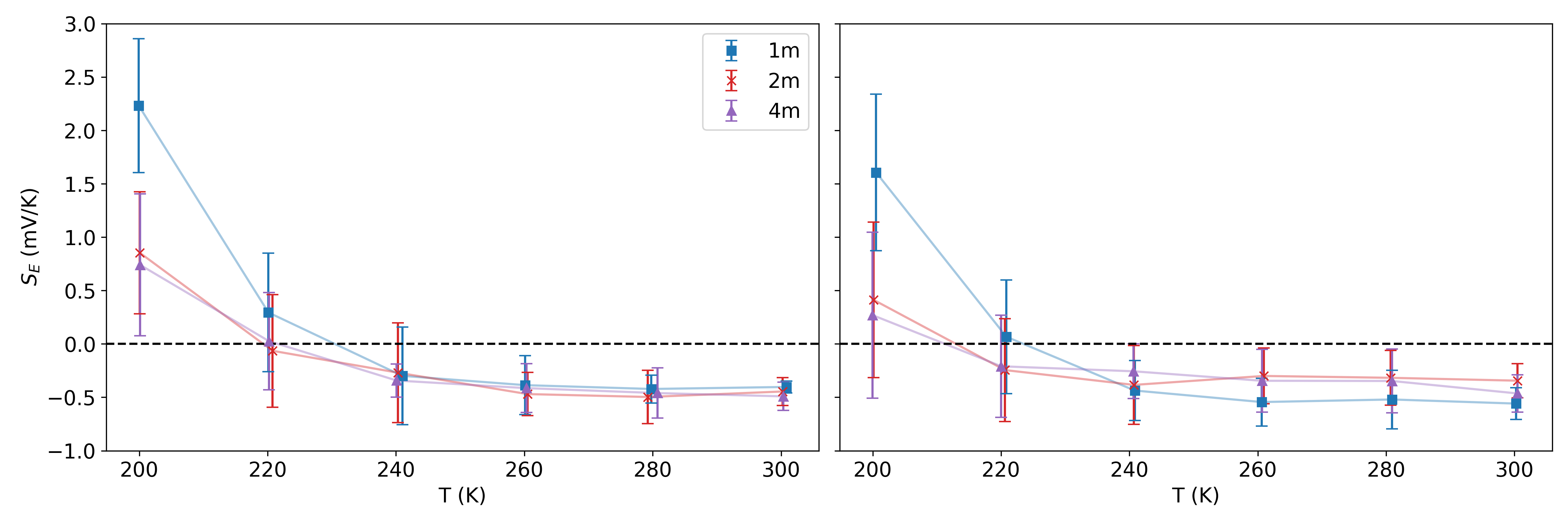}
    \caption{Seebeck coefficient of NaCl (left) and LiCl (right) solutions at 1, 2, and 4 m concentrations within the 200-300 K temperature range.}
    \label{fig:seebeck}
\end{figure*}

The relationship between water structure and thermodiffusion also applies to other transport phenomena. The Seebeck serves as an additional probe to examine how structural changes affect the thermal transport in the supercooled regime. (see Figure \ref{fig:seebeck}). Above 240 K, no significant differences are observed between NaCl and LiCl at any of the studied concentrations. Our results for 1 m NaCl at 300 K closely agree with a previous simulation study and fall within the reported error bars, despite the use of different force fields.\cite{nickel_water_2024} 
In the deeply supercooled region, we observe an inversion in the sign of the Seebeck coefficient around 220 K. Overall, our results indicate a significant Seebeck effect in the supercooled regime, with an enhancement of this effect in the deeply supercooled regime.

\section{Conclusions}

We conducted nonequilibrium molecular dynamics simulations to investigate thermal transport properties of NaCl and LiCl solutions in the supercooled water regime using TIP4P/2005 and Madrid-2019 force fields. Our results provide evidence supporting the liquid-liquid phase transition (LLPT) hypothesis in electrolyte solutions at moderate salt concentrations, \(\sim\) 1 m.

At 1 m salt concentration, the thermal conductivity shows pronounced minima near 220 K. This temperature coincides with maxima in isothermal compressibility and minima in the speed of sound. These anomalies progressively weaken with increasing concentration and disappear at 4 m, consistent with salt-induced suppression of the hypothesized LLPT\cite{perin_phase_2023}. The correlation between thermal transport anomalies and thermodynamic signatures suggests these phenomena share a common structural origin.

The Soret effect reveals complex thermodiffusion behavior with multiple sign reversals. Solutions transition from thermophobic at high temperatures (\(>\)280 K) to thermophilic upon cooling, followed by a return to thermophobic behavior in the deeply supercooled regime (\(<\)220 K). Local minima in the Soret coefficient occur between 220-250 K, with NaCl solutions displaying sharper anomalies than LiCl. These transitions correlate with structural changes in the hydrogen bonding network, as evidenced by tetrahedral order parameter analysis, which indicates an interconversion between the HDL and LDL structures as temperature changes. Notably, LiCl solutions exhibit an additional population of highly disordered local environments (\( \xi < \) 0.2), absent in NaCl, indicating that lithium ions induce stronger structural heterogeneity in the hydrogen-bond network, which may contribute to the weaker and less sharply defined thermodiffusion anomalies observed for LiCl compared to NaCl.

The Seebeck coefficients are on the order of 1 mV/K, and show sign inversions around 220-230 K, demonstrating thermoelectric sensitivity to structural transitions and confirming coupling between thermal and electrical responses.

This study establishes thermal transport measurements as sensitive probes of structural transitions in supercooled electrolyte solutions, offering new microscopic insights into water's anomalous behavior. In particular, we demonstrate a correlation between the  HDL/LDL structures and thermodiffusion. At high temperatures, when the HDL (lower tetrahedral order) structure is dominant, the solutions exhibit thermophobic responses. In contrast, they become thermophilic at temperatures where LDL, and therefore a stronger tetrahedral order, dominates.

Additional molecular simulations with different force fields and experimental validation at low temperatures and high salt concentrations would provide a route to test these theoretical predictions and advance our understanding of water's phase behavior under extreme conditions.

\section*{DATA AVAILABILITY}
The data that supports the findings of this study are available within the article.

\begin{acknowledgments}
We acknowledge the ICL RCS High-Performance Computing facility and the UK Materials and Molecular Modelling Hub for computational resources, partially funded by the EPSRC (Grant Nos. EP/P020194/1 and EP/T022213/1.
\end{acknowledgments}

\noindent
\section*{AUTHOR DECLARATIONS}
\noindent
Conflict of Interest: 
The authors have no conflicts to disclose.

\clearpage
\appendix

\section{Time Evolution and Equilibration at Low Temperatures}
\label{sec:time-evo}

\begin{figure*}[ht!]
    \centering
    \includegraphics[width=1\linewidth]{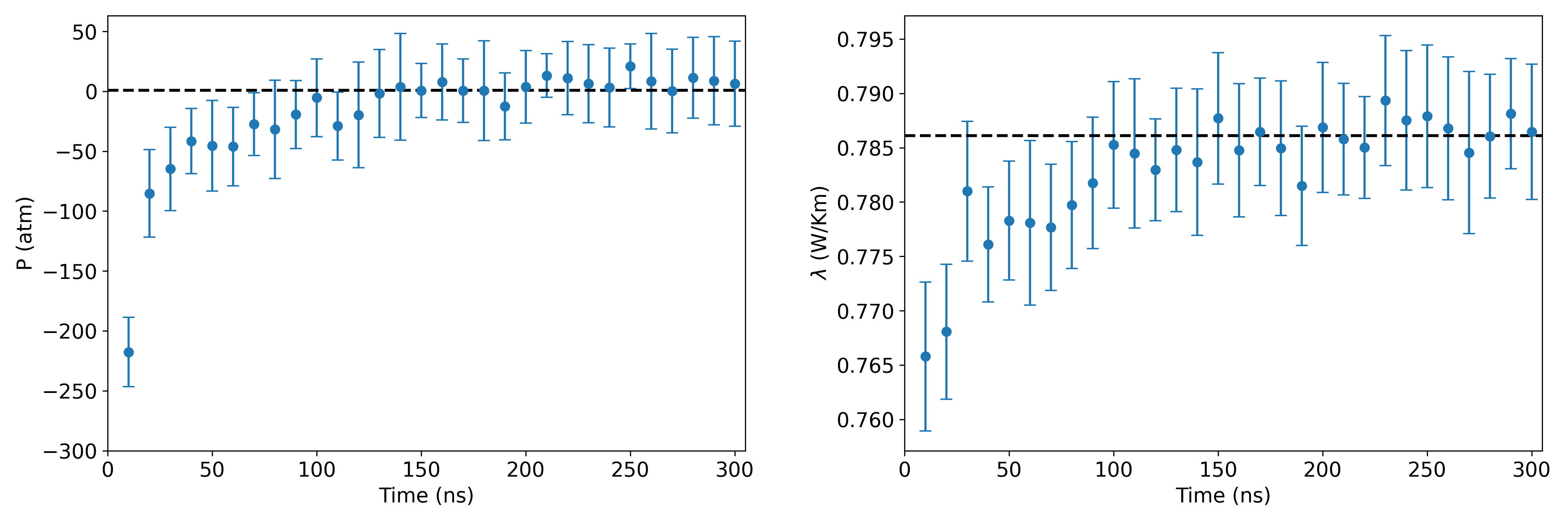}
    \caption{Time evolution of the pressure (left) and thermal conductivity \(\lambda\) (right) over 0–300 ns for the NEMD simulation of 1 M NaCl solution, with hot and cold thermostats set at 250 K and 150 K, respectively. The simulations were performed at an average temperature of 200 K and a density of 1.009 \(g/cm^3\). The dashed black lines in the left and right panels represent the target pressure (1 bar) and the calculated average thermal conductivity value, respectively.}
    \label{fig:time-evo}
\end{figure*}

Figure \ref{fig:time-evo} shows the time evolution of pressure and thermal conductivity at the lowest simulated temperature, 200 K. Similar to previous findings for supercooled pure water,\cite{zhao_thermal_2024} the alkali halide aqueous solutions display slow dynamics at low temperatures, requiring an extended period to reach equilibrium. Consequently, the initial 150 ns of the production run were excluded from the analysis.

\section{Energy Conservation Check}
\label{sec:energy_conserv}

\begin{figure}[ht!]
    \centering
    \includegraphics[width=1\linewidth]{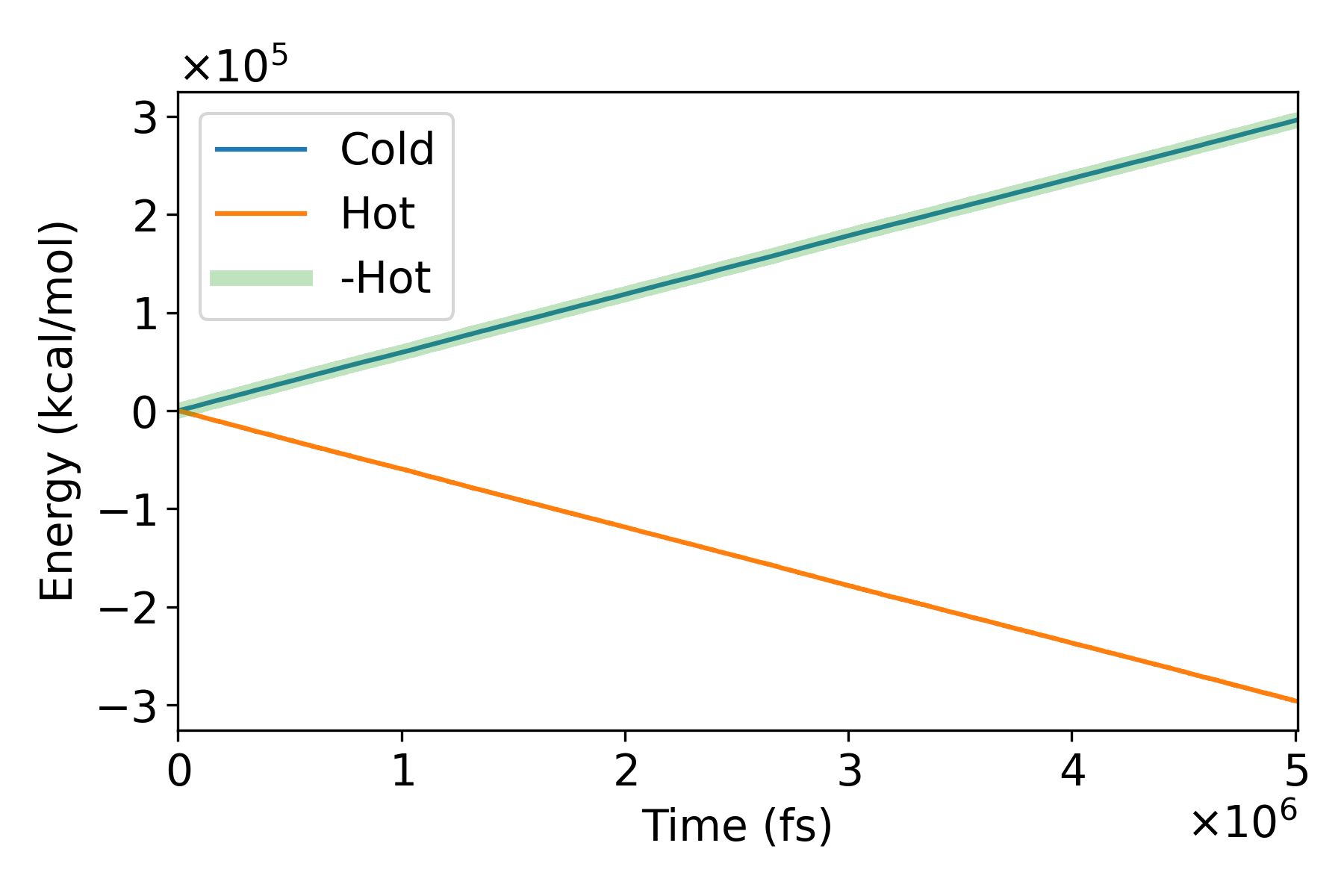}
    \caption{Energy exchange in the thermostatting regions for a representative NEMD simulation of 1 m NaCl solution at an average temperature of 280 K and a density of 1.042 \( g/cm^3\). The cold and hot thermostats were set to 330 K and 230 K, respectively. The blue and orange lines indicate the cumulative energy exchanged at the cold and hot thermostats, respectively.  For clarity in assessing energy conservation, the green line shows the negative of the energy exchanged in the hot region. The heat rate, \(\dot{Q}\), in Equation \ref{eqn: heat-flux} was obtained from the time derivative of the energy exchanged.}
    \label{fig:energy_conserv}
\end{figure}

Figure \ref{fig:energy_conserv} presents the energy exchanged between the cold and hot thermostatting regions during a representative 5 ns NEMD simulation. The consistency between the energy exchanged at the hot and cold thermostats confirms that the simulations feature good energy conservation. 

\section{Representative Local Concentration Profiles and Soret Coefficient}
\label{sec:concentration-T}

\begin{figure*}[ht!]
    \centering
    \includegraphics[width=1\linewidth]{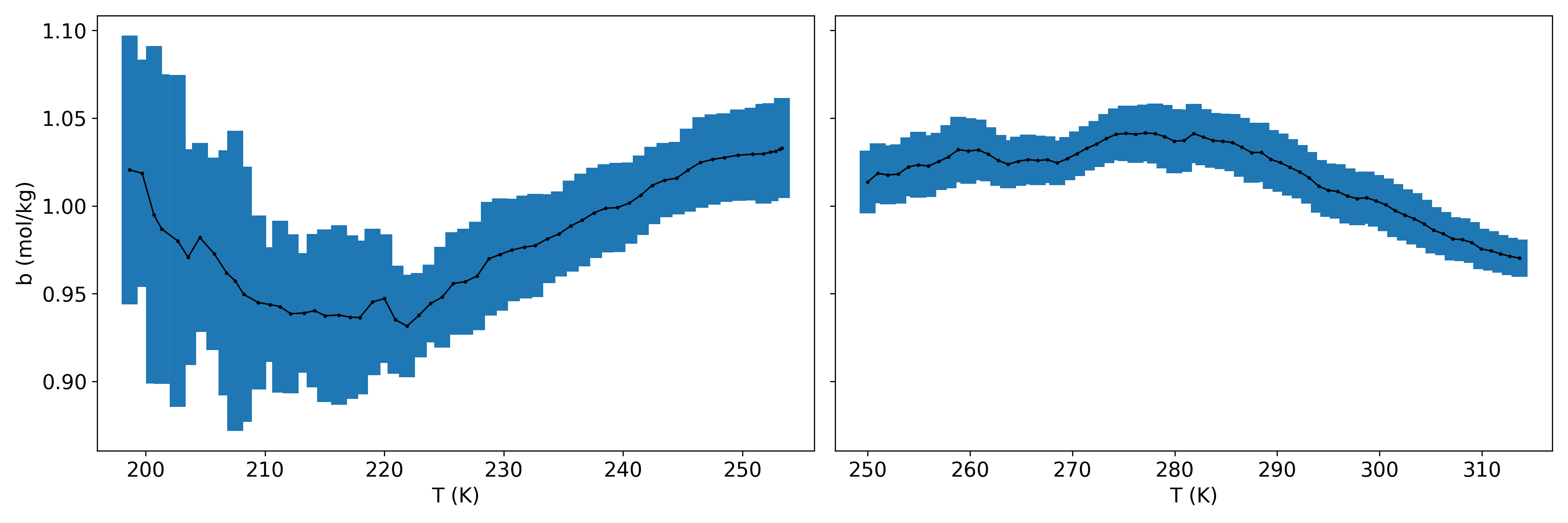}
    \caption{Local salt concentration versus temperature from representative NEMD simulations of 1 M NaCl solution over two temperature intervals: 170–270 K (left) and 230–330 K (right). The black points indicate the simulation data averaged over NEMD replicas, corresponding to average temperature and density 220 K and 1.028 \( g/cm^3\) for the left panel, and 280 K and 1.042 \( g/cm^3\) for the right panel. The blue areas around the lines represent the uncertainty in the concentration profiles.}
    \label{fig:conc_vs_T}
\end{figure*}

Figure \ref{fig:conc_vs_T} shows typical plots of local molality concentration as a function of temperature from NEMD simulations of a 1 M NaCl solution. The Soret coefficient can be obtained from the temperature derivative of these concentration profiles according to Equation \ref{eqn:Soret}. The left and right panels correspond to simulations performed at the average temperatures 220 and 280 K, respectively. The concentration features a local minimum and maximum in salt concentration around 215 and 270 K. These extrema mark the approximate temperatures where the Soret coefficient changes sign, consistent with the trend observed in Figure \ref{fig:soret} in the main text. Note that due to significant fluctuations, abrupt concentration jumps, and large uncertainties near the glass transition, particularly below 200 K, data in this low-temperature regime were excluded from the Soret coefficient calculations.

\section{Calculation of the Local Electrostatic Potential and Seebeck Coefficient}
\label{sec:Potential-T}

\begin{figure*}[ht!]
    \centering
    \includegraphics[width=1\linewidth]{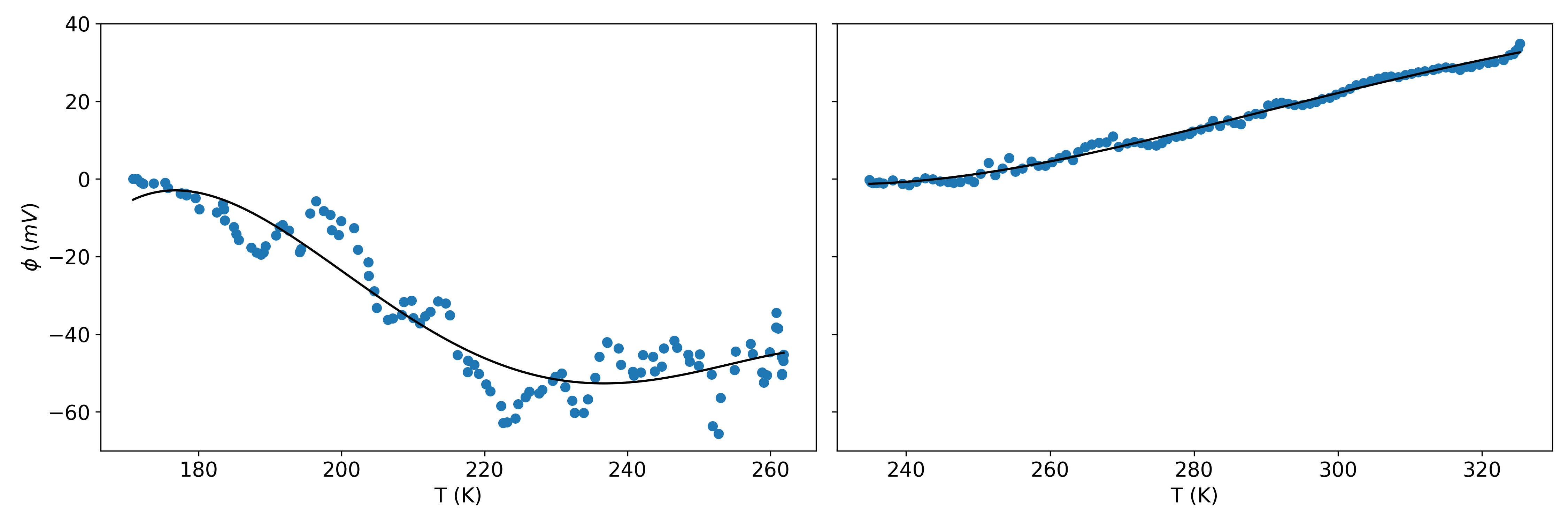}
    \caption{Electrostatic potential as a function of temperature from representative NEMD simulations of 1 M NaCl solution over the temperature ranges of 170-270 K (left) and 230-330 K (right). Blue dots indicate the NEMD simulated data, obtained at the average temperature and density of 220 K and 1.028 \( g/cm^3\) for the left panel, and 280 K and 1.042 \( g/cm^3\) for the right panel, respectively. The smoothed curve, shown as a solid black line, was obtained using a 4th-order polynomial fit, which was also used for calculating the temperature derivatives.}
    \label{fig:pot-T}
\end{figure*}

Figure \ref{fig:pot-T} shows the dependence of electrostatic potential on temperature. The local electrostatic potential \(\phi(z)\) was determined by first integrating the charge density to calculate the electric field, as described in Equation \ref{eqn: Efield}. The $z$-component of the electric field was then integrated to yield the electrostatic potential:
\begin{equation}
    \phi (z) = - \int_{0}^{z} E_{z}(z')\, dz'. 
\label{eqn:potential}
\end{equation}
\noindent
The Seebeck coefficient was evaluated using the relation $S_E(z)= -\partial \phi(z) / \partial T(z)$. We fitted the electrostatic potential as a function of temperature with a 4th-order polynomial and evaluated the negative first derivative of the fit. Both panels in Figure \ref{fig:pot-T} were calculated by analyzing trajectories spanning 5 ns, with the electrostatic potential value set to zero at the starting point of the integration.

At lower temperatures, as shown in the left panel, the electrostatic potential profiles display increased fluctuations compared to higher-temperature runs. This behavior is primarily attributed to the slow dynamics and reduced sampling efficiency in the low-temperature regime. To improve statistical reliability, the trajectory analysis was based on simulations extended to 25 ns with additional replicas. Nonetheless, the Seebeck coefficient at temperatures below 220 K still exhibits larger error bars, as shown in Figure \ref{fig:seebeck}. Despite this, a local minimum is observed around 220–230 K, indicating a sign reversal in the Seebeck effect near this temperature.

\section{Tetrahedral Order Parameter Distributions and Structural Fractionation}
\label{sec:Tetrahedral}

\begin{figure*}[ht!]
    \centering
    \includegraphics[width=1\linewidth]{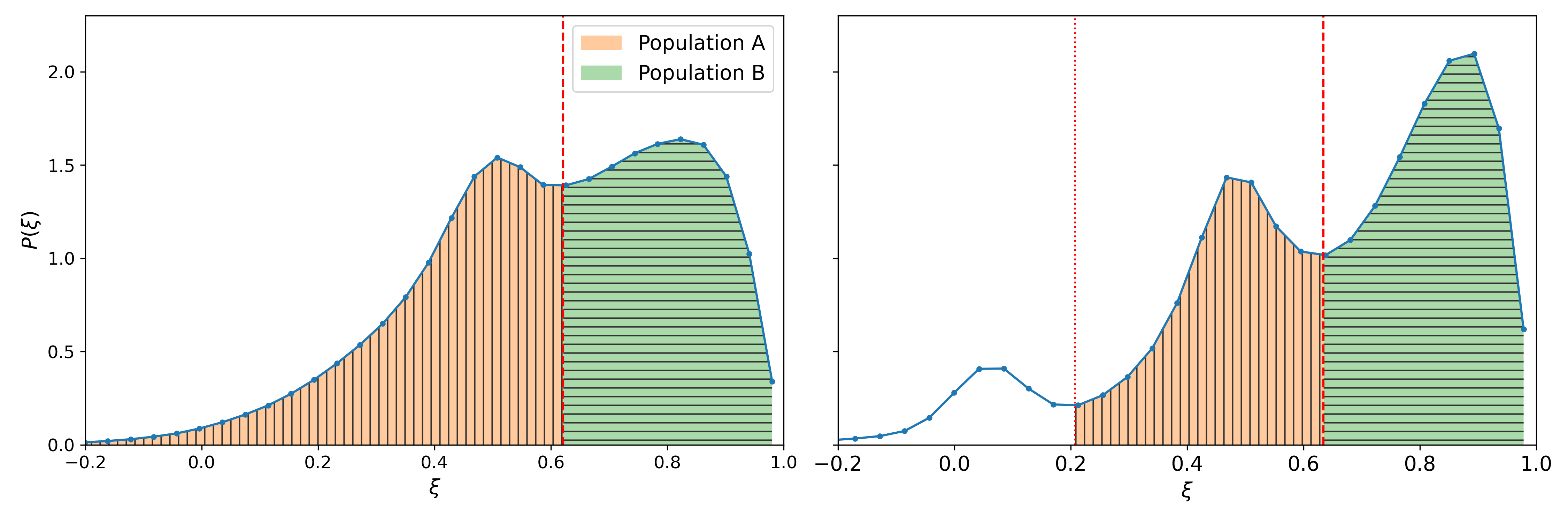}
    \caption{Representative probability density $P(\xi)$ for NaCl (left) and LiCl (right) aqueous solutions at 4 m concentration and 200 K average temperature. The two structural subpopulations are highlighted: HDL (Population A, orange with vertical hatching, less tetrahedrally ordered) and LDL (Population B, green with horizontal hatching, more tetrahedrally ordered). Red dashed lines mark the inter-peak minimum between the two main peaks in NaCl and between the middle and right peaks in LiCl; the red dotted line marks the minimum between the left and middle peaks in the LiCl solution.}
    \label{fig:HDL_LDL}
\end{figure*}

Figure \ref{fig:HDL_LDL} presents normalized representative probability densities $P(\xi)$ of the tetrahedral order parameter $\xi$, computed from NEMD trajectories using Equation \ref{eqn:order_param}. To quantify the relative populations of local structural motifs, the distributions are partitioned at the local minima between peaks, separating the less tetrahedrally coordinated, high-density liquid-like (HDL) and the more tetrahedrally ordered, low-density liquid-like (LDL) regions. These minima serve as integration boundaries for estimating the areas associated with each population.

In LiCl solutions, an additional peak is observed at low $\xi$ values ($\xi \lesssim 0.2$), corresponding to highly disordered local structures. This feature is excluded from consideration in the HDL and LDL population analysis. In such cases, the HDL region is defined between the first and second minima, while the LDL region lies beyond the second minimum. The fractions of HDL and LDL structures are obtained by integrating their respective regions and normalizing by the combined area.

\section{NEMD Simulation Results}
\label{sec:nemd-table}

The results of the NEMD simulations are given in Table \ref{table:NEMD-data}.

\newcommand{\LTHeaderNEMD}{%
T & T$_{\text{hot}}$ & T$_{\text{cold}}$ & $\rho$ & $P$ & $\lambda$ & $S_E$ \\
{[K]} & {[K]} & {[K]} & {[g/cm$^3$]} & {[atm]} & {[W\,m$^{-1}$\,K$^{-1}$]} & {[mV\,K$^{-1}$]} \\
}
{\setlength{\LTcapwidth}{0.9\textwidth}
\begin{longtable}{
  C{1.5cm}  
  C{1.5cm}  
  C{1.5cm}  
  C{1.5cm}  
  C{2cm}  
  C{3cm}  
  C{3cm}  
}
\toprule
\LTHeaderNEMD
\midrule
\endfirsthead

\caption[]{(continued)}\\
\toprule
\LTHeaderNEMD
\midrule
\endhead

\bottomrule
\endlastfoot
\multicolumn{7}{c}{1m NaCl} \\
\midrule
200.2 & 249.9 & 150.0 & 1.009 & 8.9(22.2) & 0.786(0.016) & 2.233(0.627) \\
220.3 & 270.1 & 170.1 & 1.028 & -4.2(9.7) & 0.777(0.013) & 0.294(0.555) \\
239.9 & 291.0 & 190.0 & 1.041 & 12.5(3.1) & 0.792(0.010) & -0.299(0.458) \\
260.2 & 310.1 & 210.1 & 1.044 & 10.6(3.9) & 0.808(0.009) & -0.386(0.276) \\
280.0 & 329.8 & 229.9 & 1.042 & 4.6(1.8)  & 0.826(0.009) & -0.422(0.131) \\
300.0 & 350.8 & 249.9 & 1.036 & 6.5(2.3)  & 0.840(0.004) & -0.403(0.056) \\
\midrule
\multicolumn{7}{c}{2m NaCl} \\
\midrule
200.1 & 250.1 & 150.0 & 1.066 & 10.2(25.5) & 0.786(0.009) & 0.853(0.573) \\
220.4 & 270.8 & 170.0 & 1.078 & 0.3(7.6)  & 0.787(0.008) & -0.065(0.527) \\
240.2 & 290.3 & 189.9 & 1.086 & 4.9(4.7)  & 0.791(0.006) & -0.270(0.468) \\
260.2 & 310.5 & 210.0 & 1.085 & 3.9(3.7)  & 0.803(0.007) & -0.470(0.202) \\
280.1 & 329.2 & 229.9 & 1.080 & -1.5(1.2) & 0.815(0.008) & -0.497(0.251) \\
300.0 & 350.3 & 249.9 & 1.067 & 8.5(1.8)  & 0.825(0.001) & -0.446(0.131) \\
\midrule
\multicolumn{7}{c}{4m NaCl} \\
\midrule
200.0 & 250.1 & 149.9 & 1.166 & -88.6(34.0) & 0.767(0.014) & 0.741(0.665) \\
220.1 & 270.2 & 170.1 & 1.169 & -72.0(7.1)  & 0.771(0.015) & 0.025(0.456) \\
240.1 & 290.1 & 189.9 & 1.166 & -63.2(4.9)  & 0.778(0.007) & -0.343(0.154) \\
260.1 & 310.3 & 209.7 & 1.158 & -66.9(3.0)  & 0.786(0.007) & -0.414(0.230) \\
280.1 & 330.8 & 229.8 & 1.147 & -66.2(1.0)  & 0.792(0.007) & -0.461(0.235) \\
299.8 & 350.3 & 249.9 & 1.134 & -33.4(2.2)  & 0.801(0.004) & -0.491(0.133) \\
\midrule
\multicolumn{7}{c}{1m LiCl} \\
\midrule
200.3 & 250.5 & 149.9 & 0.985 & 38.7(16.1) & 0.785(0.011) & 1.606(0.733) \\
220.6 & 270.8 & 169.9 & 1.005 & 7.0(16.8)  & 0.775(0.010) & 0.067(0.533) \\
240.3 & 290.8 & 190.1 & 1.015 & -7.9(4.3)  & 0.783(0.008) & -0.437(0.281) \\
260.3 & 310.6 & 210.1 & 1.022 & -2.3(2.3)  & 0.802(0.005) & -0.545(0.224) \\
280.2 & 330.9 & 229.9 & 1.023 & 1.1(1.5)   & 0.814(0.005) & -0.521(0.275) \\
300.1 & 350.2 & 249.9 & 1.018 & 1.8(0.8)   & 0.835(0.004) & -0.559(0.149) \\
\midrule
\multicolumn{7}{c}{2m LiCl} \\
\midrule
200.0 & 250.1 & 149.6 & 1.019 & 76.2(17.8)  & 0.780(0.009) & 0.412(0.728) \\
220.5 & 270.6 & 169.9 & 1.036 & -12.8(18.6) & 0.780(0.008) & -0.245(0.482) \\
240.2 & 290.7 & 190.1 & 1.042 & -9.4(2.7)   & 0.781(0.006) & -0.384(0.369) \\
260.2 & 311.0 & 210.0 & 1.046 & -5.3(3.5)   & 0.794(0.006) & -0.300(0.262) \\
280.2 & 330.7 & 230.2 & 1.044 & -1.7(1.0)   & 0.808(0.004) & -0.318(0.255) \\
300.1 & 350.5 & 249.9 & 1.039 & 4.3(1.1)    & 0.821(0.003) & -0.345(0.159) \\
\midrule
\multicolumn{7}{c}{4m LiCl} \\
\midrule
200.0 & 250.0 & 149.5 & 1.082 & -49.8(37.6) & 0.760(0.010) & 0.269(0.777) \\
220.1 & 270.0 & 169.9 & 1.088 &  -7.3(14.4) & 0.762(0.007) & -0.210(0.479) \\
240.2 & 290.6 & 190.2 & 1.089 &  -3.5(4.5)  & 0.769(0.006) & -0.256(0.257) \\
260.1 & 310.6 & 209.7 & 1.087 &  -0.2(2.1)  & 0.778(0.007) & -0.345(0.294) \\
280.1 & 330.9 & 229.9 & 1.084 &   0.4(0.9)  & 0.784(0.006) & -0.347(0.299) \\
300.0 & 350.5 & 250.0 & 1.078 &  -2.2(0.9)  & 0.796(0.003) & -0.464(0.176) \\
\midrule
\caption{NEMD simulation results for NaCl and LiCl aqueous solutions at 1, 2, and 4 m. Columns list the mean temperature $T$ of the simulation cell, the mean temperatures of the hot and cold reservoirs $T_{\mathrm{hot}}$ and $T_{\mathrm{cold}}$, density $\rho$, pressure $P$, thermal conductivity $\lambda$, and Seebeck coefficient $S_E$. Values are time averages over steady-state windows, and numbers in parentheses denote standard deviations across independent replicas.}

\label{table:NEMD-data}
\end{longtable}
}

\section{Equilibrium NPT Simulation Results}
\label{sec:NPT-table}
The Equilibrium NPT simulation results are presented in Table \ref{table:NPT-data}.
\clearpage
{\setlength{\LTcapwidth}{0.9\textwidth}
\DeclareRobustCommand{\cation}{\mathrm{Li}}
\newcommand{\LTHeaderLi}{%
T & $\rho$ & $\alpha$ & $\kappa_T$ & $c_s$ &
n$_{\cation\text{--}O}$ & $r_{\cation\text{--}O}$ &
n$_{\text{Cl--O}}$ & $r_{\text{Cl--O}}$ \\
{[K]} & {[g/cm$^3$]} & {[$\times 10^{-4}$ K$^{-1}$]} &
{[$\times 10^{-10}$ m$^3$J$^{-1}$]} & {[m/s]} &
& {[nm]} & & {[nm]} \\
}
\begin{longtable}{
  C{1.0cm}  
  C{1.2cm}  
  C{2.7cm}  
  C{2.7cm}  
  C{2.2cm}  
  C{1cm}  
  C{1cm}  
  C{1cm}  
  C{1cm}  
}
\toprule
T & $\rho$ & $\alpha$ & $\kappa_T$ & $c_s$ &
n$_{\text{Na--O}}$ & $r_{\text{Na--O}}$ &
n$_{\text{Cl--O}}$ & $r_{\text{Cl--O}}$ \\
{[K]} & {[g/cm$^3$]} & {[$\times 10^{-4}$ K$^{-1}$]} &
{[$\times 10^{-10}$ m$^3$J$^{-1}$]} & {[m/s]} &
& {[nm]} & & {[nm]} \\
\midrule
\endfirsthead

\caption[]{(continued)} \\
\toprule
\endhead

\bottomrule
\endlastfoot

\multicolumn{9}{c}{1m NaCl} \\
\midrule
200 & 1.009 & -2.54(0.58) & 4.11(0.29) & 1565.4(36.0) & 5.68 & 0.31 & 5.76 & 0.36 \\
220 & 1.028 & -6.34(0.26) & 5.79(0.25) & 1301.5(31.4) & 5.67 & 0.31 & 5.78 & 0.36 \\
240 & 1.041 & -3.96(0.14) & 5.05(0.19) & 1391.1(19.4) & 5.63 & 0.31 & 5.79 & 0.36 \\
260 & 1.044 &  0.79(0.18) & 4.50(0.05) & 1476.3(10.3) & 5.61 & 0.32 & 5.80 & 0.36 \\
280 & 1.042 &  2.15(0.13) & 4.24(0.07) & 1518.2(10.9) & 5.57 & 0.32 & 5.80 & 0.36 \\
300 & 1.036 &  3.71(0.16) & 4.11(0.06) & 1556.4(9.4)  & 5.50 & 0.31 & 5.80 & 0.36 \\
\midrule
\multicolumn{9}{c}{2m NaCl} \\
\midrule
200 & 1.066 & -8.04(0.56) & 4.85(0.37) & 1393.3(27.3) & 5.78 & 0.32 & 5.82 & 0.36 \\
210 & 1.070 & -5.76(0.51) & 5.40(0.27) & 1335.6(27.6) & 5.75 & 0.32 & 5.82 & 0.36 \\
220 & 1.078 & -5.29(0.32) & 4.83(0.22) & 1386.9(17.1) & 5.73 & 0.31 & 5.82 & 0.36 \\
240 & 1.086 & -1.50(0.20) & 4.17(0.16) & 1486.1(9.9)  & 5.68 & 0.32 & 5.82 & 0.36 \\
260 & 1.085 &  1.83(0.16) & 3.95(0.14) & 1532.8(10.5) & 5.62 & 0.32 & 5.89 & 0.36 \\
280 & 1.080 &  3.28(0.07) & 3.72(0.04) & 1581.6(9.2)  & 5.56 & 0.32 & 5.90 & 0.36 \\
300 & 1.067 &  4.27(0.07) & 3.72(0.04) & 1614.5(9.8)  & 5.51 & 0.32 & 5.93 & 0.37 \\
\midrule
\multicolumn{9}{c}{4m NaCl} \\
\midrule
200 & 1.166 & -2.87(0.65) & 4.16(0.26) & 1437.5(27.9) & 5.85 & 0.32 & 5.87 & 0.36 \\
220 & 1.169 & -0.64(0.32) & 3.85(0.20) & 1515.4(29.3) & 5.78 & 0.32 & 5.89 & 0.36 \\
240 & 1.166 &  1.67(0.27) & 3.40(0.13) & 1607.7(20.5) & 5.70 & 0.32 & 5.90 & 0.36 \\
260 & 1.158 &  2.95(0.19) & 3.18(0.03) & 1663.9(12.8) & 5.61 & 0.32 & 5.91 & 0.36 \\
280 & 1.147 &  4.01(0.09) & 3.15(0.03) & 1689.7(10.7) & 5.51 & 0.31 & 5.93 & 0.36 \\
300 & 1.134 &  4.77(0.09) & 3.14(0.01) & 1707.8(7.1)  & 5.47 & 0.32 & 5.94 & 0.36 \\
\midrule
\newpage        
\LTHeaderLi
\midrule
\multicolumn{9}{c}{1m LiCl} \\
\midrule
200 & 0.985 & -6.33(0.44) & 4.51(0.42) & 1503.0(42.4) & 4.00 & 0.24 & 5.62 & 0.36 \\
220 & 1.005 & -7.32(0.25) & 5.02(0.20) & 1411.0(38.9) & 4.00 & 0.24 & 5.71 & 0.36 \\
240 & 1.015 & -3.89(0.15) & 5.01(0.15) & 1428.6(30.1) & 4.00 & 0.25 & 5.71 & 0.36 \\
260 & 1.022 & -1.72(0.15) & 4.43(0.07) & 1493.2(14.9) & 4.00 & 0.26 & 5.76 & 0.36 \\
280 & 1.023 &  1.56(0.29) & 4.31(0.05) & 1524.4(17.9) & 4.00 & 0.26 & 5.82 & 0.37 \\
300 & 1.018 &  3.50(0.19) & 4.19(0.05) & 1540.5(12.5) & 4.00 & 0.26 & 5.85 & 0.37 \\
\midrule
\multicolumn{9}{c}{2m LiCl} \\
\midrule
200 & 1.019 & -6.50(0.37) & 4.01(0.20) & 1563.4(31.3) & 4.00 & 0.23 & 5.64 & 0.36 \\
220 & 1.036 & -5.19(0.45) & 4.19(0.19) & 1549.7(20.1) & 4.00 & 0.24 & 5.75 & 0.36 \\
240 & 1.042 & -3.04(0.18) & 4.02(0.11) & 1551.7(17.1) & 4.00 & 0.25 & 5.75 & 0.36 \\
260 & 1.046 & -0.68(0.18) & 3.93(0.07) & 1569.7(12.4) & 4.00 & 0.25 & 5.76 & 0.36 \\
280 & 1.044 &  2.19(0.08) & 3.84(0.07) & 1584.8(12.0) & 4.00 & 0.26 & 5.83 & 0.37 \\
300 & 1.039 &  3.79(0.12) & 3.78(0.06) & 1618.3(12.1) & 4.00 & 0.28 & 5.90 & 0.37 \\
\midrule
\multicolumn{9}{c}{4m LiCl} \\
\midrule
200 & 1.082 & -2.63(0.51) & 3.61(0.18) & 1597.6(31.5) & 4.00 & 0.24 & 5.74 & 0.36 \\
220 & 1.088 & -0.75(0.43) & 3.48(0.14) & 1634.4(24.7) & 4.00 & 0.24 & 5.73 & 0.36 \\
240 & 1.089 & -0.24(0.37) & 3.26(0.09) & 1678.7(15.8) & 4.00 & 0.25 & 5.74 & 0.36 \\
260 & 1.087 &  0.75(0.12) & 3.24(0.06) & 1694.6(12.0) & 4.00 & 0.25 & 5.79 & 0.36 \\
280 & 1.084 &  2.55(0.14) & 3.18(0.05) & 1715.3(11.5) & 4.00 & 0.26 & 5.81 & 0.36 \\
300 & 1.078 &  4.03(0.11) & 3.14(0.06) & 1728.4(11.6) & 4.00 & 0.26 & 5.89 & 0.37 \\
\midrule
\caption{Thermophysical properties (density $\rho$, isobaric thermal expansion coefficient $\alpha$, isothermal compressibility $\kappa_T$, adiabatic speed of sound $c_s$) and structural properties (coordination numbers between the ions and oxygen in water, with cutoff radius $r$ at the first minimum of the ion--O $g(r)$) for NaCl and LiCl aqueous solutions at 1, 2, and 4 m over 200--300 K from equilibrium NPT simulations. Values are means and uncertainties in parentheses are standard deviations across independent replicas.}

\label{table:NPT-data}
\end{longtable}
}

\clearpage

\bibliographystyle{aipnum4-1}
\bibliography{mybibliography.bib}

\end{document}